\documentclass[10pt,letterpaper]{article}
\usepackage[top=0.85in,left=2.75in,footskip=0.75in,marginparwidth=2in]{geometry}

\usepackage[utf8]{inputenc}

\usepackage{cite}

\usepackage{nameref,hyperref}

\usepackage{microtype}
\DisableLigatures[f]{encoding = *, family = * }

\raggedright
\usepackage{changepage}

\usepackage[aboveskip=1pt,labelfont=bf,labelsep=period,singlelinecheck=off]{caption}

\makeatletter
\renewcommand{\@biblabel}[1]{\quad#1.}
\makeatother

\usepackage{lastpage,fancyhdr,graphicx}
\usepackage{epstopdf}
\fancyheadoffset[L]{2.25in}
\fancyfootoffset[L]{2.25in}

\usepackage{color}

\definecolor{Gray}{gray}{.25}

\usepackage{graphicx}

\usepackage{sidecap}

\usepackage{wrapfig}
\usepackage[pscoord]{eso-pic}
\usepackage[fulladjust]{marginnote}
\reversemarginpar

\begin{document}
\vspace*{0.35in}

\begin{flushleft}
{\Large
\textbf\newline{Field-free all-optical switching and electrical read-out \\ of Tb/Co-based magnetic tunnel junctions.}
}
\newline
\\

D. Salomoni\textsuperscript{1},
Y. Peng\textsuperscript{2},
L. Farcis\textsuperscript{1},
S. Auffret\textsuperscript{1},
M. Hehn\textsuperscript{2},
G. Malinowski\textsuperscript{2},
S. Mangin\textsuperscript{2},
B. Dieny\textsuperscript{1},
L. D. Buda-Prejbeanu\textsuperscript{1},
R.C. Sousa\textsuperscript{1},
I. L. Prejbeanu\textsuperscript{1,*},
\\
\bigskip
\bf{1} Univ Grenoble Alpes, CEA, CNRS, Grenoble INP, SPINTEC, 38000 Grenoble, France.
\\
\bf{2} Institut Jean Lamour, UMR CNRS 7198, Université de Lorraine, 54011 Nancy, France.
\\
\bigskip
* ioan-lucian.prejbeanu@cea.fr

\end{flushleft}
\begin{flushleft}
\section{Abstract}
\end{flushleft}
Switching of magnetic tunnel junction using femto-second laser enables a possible path for THz frequency memory operation, which means writing speeds 2 orders of magnitude faster than alternative electrical approaches based on spin transfer or spin orbit torque. In this work we demonstrate successful field-free 50fs single laser pulse driven magnetization reversal of [Tb/Co] based storage layer in a perpendicular magnetic tunnel junction. The nanofabricated magnetic tunnel junction devices have an optimized bottom reference electrode and show Tunnel Magnetoresistance Ratio values (TMR) up to 74\% after patterning down to sub-100nm lateral dimensions.  Experiments on continuous films reveal peculiar reversal patterns of concentric rings with opposite magnetic directions, above certain threshold fluence. These rings have been correlated to patterned device switching probability as a function of the applied laser fluence. Moreover, the magnetization reversal is independent on the duration of the laser pulse. According to our macrospin model, the underlying magnetization reversal mechanism can be attributed to an in-plane reorientation of the magnetization due to a fast reduction of the out-of-plane uniaxial anisotropy. These aspects are of great interest both for the physical understanding of the switching phenomenon and their consequences for all-optical-switching memory devices, since they allow for a large fluence operation window with high resilience to pulse length variability.


\section*{Introduction}

Conventional computer memories have evolved into a many-level hierarchy where the operation speed, storage density and cost, define a trade-off leading to the implementation of different memory technologies. As the physical limits of complementary metal oxide semiconductor (CMOS) memories  are reached, the possibility to replace volatile memory with fast non-volatile alternatives has been a compelling argument of Magnetoresistive Random Access Memory (MRAM)\cite{DIE20, KIM15, KIM19}. Perpendicular spin transfer torque (STT) MRAM, due to advantages such as reliable switching, low energy consumption and easy integration with CMOS technology \cite{DIE20}, is the most widely foundry adopted spintronics based memory \cite{SOL96, BER96, RAL08}. High-performance STT-MTJ cells are among the first MRAMs to have recently been commercialized as embedded flash memory and last-level cache replacements. The FeCoB/MgO/FeCoB system has become the basis of most magnetic tunnel junction (MTJ) designs, due to its high tunnel magnetoresistance effect (TMR) \cite{YUA04, PAR04}, improved readability and scaling of interfacial perpendicular magnetic anisotropy (PMA) \cite{BRA12, IKE10} based devices. Moreover, in STT cells the reading and writing paths are shared enabling a compact design \cite{AND14, KHV13}, however fast switching in the precessional regime requires switching voltages close to the barrier breakdown voltage.

Alternative Spin-Orbit Torque (SOT) technology, is more suitable for high-speed and low error rate operation \cite{MAN19, MIR11}. The SOT three terminal devices with separate reading and writing paths enable better endurance combined with fast switching, creating a potential replacement for SRAM \cite{KRI22}. Its main drawback is the larger bit-cell area, and although switching pulses as short as 100ps are possible, the application of a field is generally required to obtain deterministic switching which complexifies the cell design. 
The all-optical-switching (AOS) technology envisions to further accelerate the magnetization reversal process by enabling writing the cells at femtosecond time scales. Single Pulse All Optical Helicity Independent Switching (AO-HIS) is a phenomenon by which the magnetization of a nanostructure can be reversed by a single femtosecond laser pulse. The method is ultrafast and does not require use of any applied field. 
Since its discovery in GdFeCo ferrimagnetic systems \cite{STA07}, single pulse AO-HIS had been limited to Gd-based alloys or Gd/FM bilayers, where FM is a ferromagnetic layer \cite{ RAD11, OST12}. Only recently few other material systems have shown switching driven by ultrashort laser pulses: MnRuGa ferrimagnetic Heusler alloys \cite{VED20}, Tb/Co multilayers \cite{AVI19, AVI20}, Tb/Fe and Tb$_{32}$Co$_{68}$/Co trilayer\cite{PEN23}. As demonstrated in this paper, AO-HIS mechanism for Tb based heterostructures is very different than the one observed for Gd based materials.

Using AOS as a write mechanism in MRAM is expected to enable writing at speeds 2 orders of magnitude faster than electrical alternatives with an energy as low as 16fJ/bit  \cite{KIM19},  while maintaining non-volatility \cite{KIM19, VED20}. This, not only increases speed and reduces power consumption, but also allows for more compact two terminal design of STT devices, with no compromise in endurance. 

The first successful AOS operation of a micron-sized p-MTJ cell was demonstrated by Chen et al \cite{CHE17}, using a GdFeCo alloy storage layer, leading to a small TMR ratio of 0.6\%. It was later improved by Wang et al.\cite{WAN22}, who reported switching of Co/Gd bilayer tunnel junctions with cell sizes down to 3$\mu$m lateral size and TMR ratio of 34.7\%. The increase of TMR was possible thanks to a clever design demonstrated by  Avilés‑Félix et al.\cite{AVI19, AVI20} where to enable AOS in a MTJ the optically-switchable material is coupled magnetically to the FeCoB interface of the storage layer. 
The initial reports of a storage layer based on [Tb/Co]$_{N}$ multilayer coupled to a FeCoB showed the possibility to adjust the Co/Tb composition in order to control the effective perpendicular anisotropy of magnetic tunnel junctions \cite{ AVI19, AVI20}. The perpendicular anisotropy could be maintained without degradation even after annealing at $250^{o}C$ to obtain  38\% TMR after patterning. Thin film experiments of the same storage layer stacks confirmed AO-HIS of the magnetization with both 60fs and 5ps laser pulses for incident fluences down to 3.5$mJ/cm^{2}$. However, AOS demonstration on patterned devices and a clear understanding of the switching mechanism of this system is missing so far.

In this work, we developed MTJ devices comprising a [Tb/Co]$_{5}$ based top storage layer with diameters from 300nm down to 80nm exhibiting TMR values up to 74\%. We demonstrate field-free helicity independent all optical toggle switching driven by 50fs single light pulse on a [Tb/Co] based p-MTJ patterned to sub-100nm lateral dimensions. The comparison of patterned device properties to those of continuous film, shows that high anisotropy of the Tb/Co based electrode is maintained after patterning. The integration of a stable reference layer, with low offset field, does not degrade the AOS stack properties. Furthermore, the AOS behaviour seen at film level correlates with pattern device switching probability, despite the evident changes in the magnetostatic interaction. The underlying reversal mechanism that takes place in these systems is analyzed in the framework of a macrospin model accounting for the temperature dependence of the anisotropy. 

\section{\label{sec:level2}Results and discussion:}

\begin{wrapfigure}[23]{l}{70mm}
\includegraphics[width=75mm]{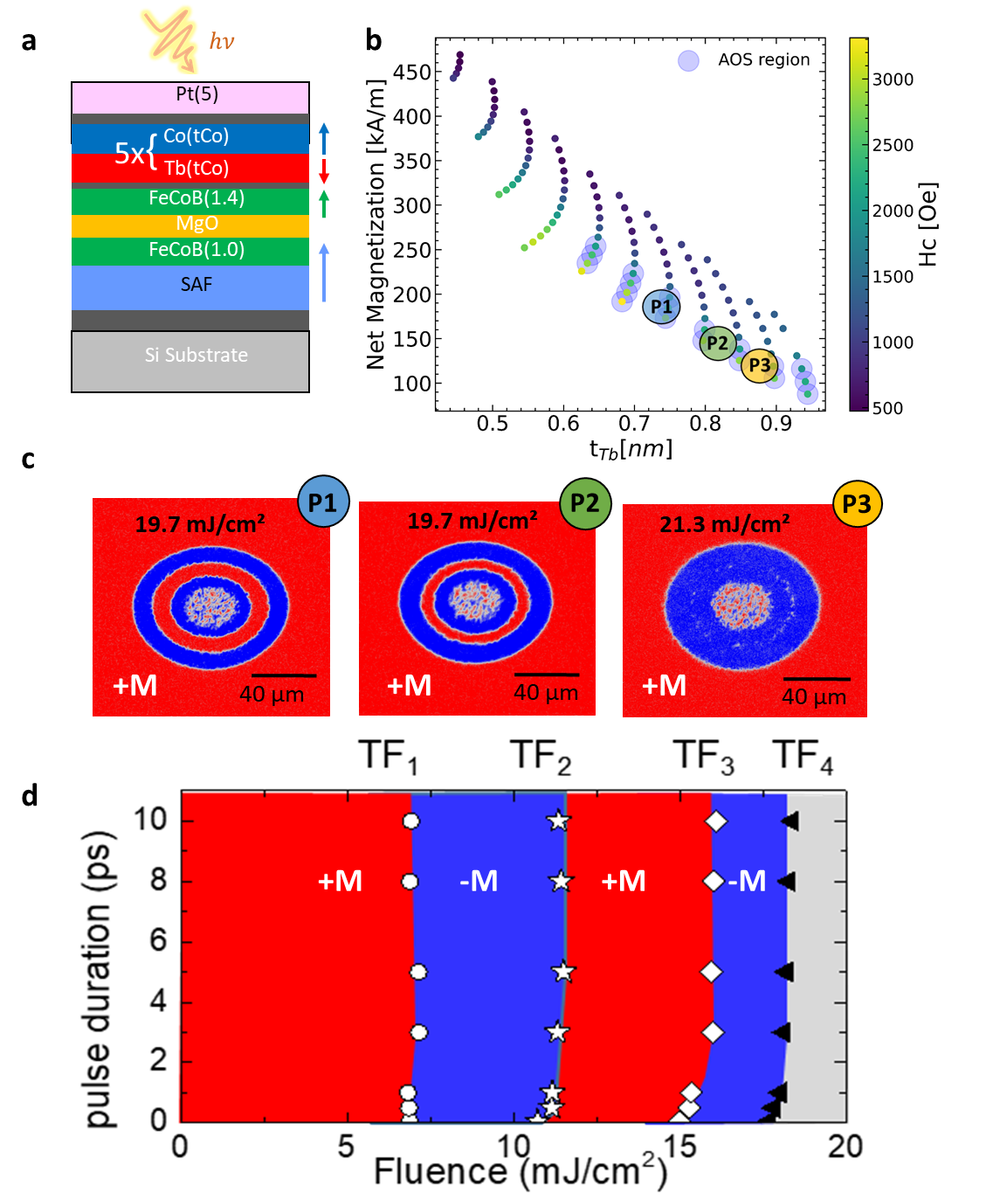}
\captionsetup{labelformat=empty} 
\caption{} 
\label{fig:stack} 
\end{wrapfigure}

Our samples depicted in Fig.~\ref{fig:stack}a have a bottom reference stack composed of a synthetic antiferromagnet based [Co/Pt] multilayer, as the one used in conventional STT-MRAM cells. The complete magnetic stack deposited by DC magnetron sputtering is: Ta(3)/Pt(25)/[Co(0.5)/Pt(0.25)]x5/Co(0.5)/Ru(0.9)/ [Co(0.5)/Pt(0.25)]x2/Co(0.5)/W(0.3)/FeCoB(1)/ MgO(t$_{Mg}$)/FeCoB(1.4)/W(0.2)/[Tb(t$_{Tb}$)/Co(t$_{Co}$)]x5/ W(2)/Pt(5), where all the thicknesses in brackets are expressed in nm. Samples were deposited using Ar pressure of $2 \times 10^{-3}$ mbar, base pressure of $10^{-8}$ mbar and annealed at $250^{o}C$. As shown in Fig.~\ref{fig:stack}b the samples were deposited using a crossed-wedge thicknesses for Tb and either MgO or Co to allow for the investigation of devices properties as function of layer thickness.

\marginpar{
\vspace{.7cm} 
\color{Gray} 
\textbf{Figure \ref{fig:stack}} 
a) AOS-MTJ stack with bottom-pinned and [Tb/Co]ML storage layer schematic. b) Net magnetization vs Tb layer thickness extracted from MOKE measurements, with highlighted AOS region, the second thickness wedge variation is $t_{Co}$ from 1.0 to 1.8nm. c) Background subtracted MOKE images after single 50fs laser pulse of different fluences,  and  for positions: P1 ($t_{Co}$:1.37nm,$t_{Tb}$:0.76nm), P2 ($t_{Co}$:1.41nm,$t_{Tb}$:0.80nm) and P3($t_{Co}$:1.45nm,$t_{Tb}$:0.86nm). d) Pulse duration vs fluence experimental state diagram of the magnetization reversal. Threshold boundaries: TF1 ring outer-border; TF2 first ring; TF3 the second ring; TF4 center demagnetized multidomain.
}

\begin{wrapfigure}[19]{l}{74mm}
\includegraphics[width=75mm]{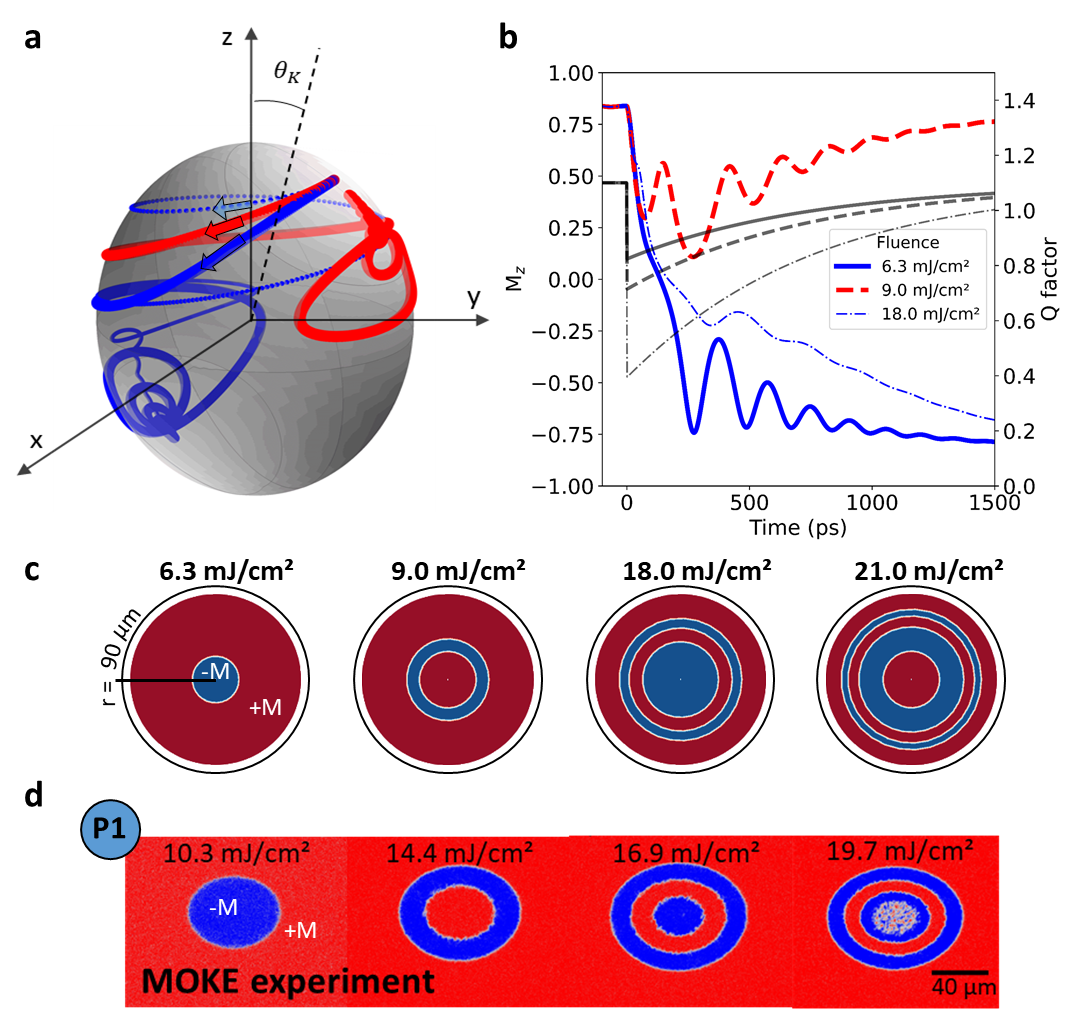}
\captionsetup{labelformat=empty} 
\caption{} 
\label{fig:LLG} 
\end{wrapfigure}

The AO-HIS properties of our multilayered samples before nanopatterning were systematically analysed by magneto-optical Kerr effect  microscopy (MOKE). The pump laser beam used in the AOS measurement was a Ti:sapphire femtosecond-laser source with wavelength and repetition rate of 800nm and 5kHz, respectively. The Gaussian beam diameter of $100\mu m$ was determined by directly observing the beam at the focal plane of the microscope lens and later confirmed using the reversed domain size as function of the pulse energy. Transmission configuration was used for samples grown on glass substrates, where the pump laser excitation was applied on one side of the sample and MOKE microscopy observations done on the substrate side. For samples grown on silicon wafers, a reflection configuration was used where the pump and probe lights are both on the sample side. Regardless of the configuration, the system shows AOS close to the magnetization compensation region at room temperature and on the Co-rich region where the Co magnetization is dominant. Similar to previous reported [Tb/Co] multilayers [\cite{AVI20}] with 5 bilayer repetitions annealed at $250^{o}C$ the Tb thickness range is from 0.6 to 0.9nm, and respectively 1.2 to 1.5nm for Co. The probe source used for the background subtracted MOKE images, shown in Fig.~\ref{fig:stack}c  and Fig.~\ref{fig:LLG}d, is a LED light with a wavelength of 628nm. After each single laser pulse a complete full signal amplitude toggle reversal of the storage layer occurs. This behavior is reproducible for the storage layer stack up to 150,000 consecutive pulses, suggesting that the toggle switching has a large endurance and repeatability for the above thickness ranges.

For fluences higher than 11mJ/cm$^{2}$, the reversed domains exhibit concentric rings with opposite magnetic directions (as seen in Fig.~\ref{fig:stack}c and Fig.~\ref{fig:LLG}d). It is worth mentioning that similar behaviour was also observed for samples with only [Tb/Co] multilayers, excluding any contribution from the SAF reference layer to the magnetization reversal, except for its impact as a heat sink structure increasing the fluence threshold of the different regions \cite{PEN23}. Based on the MOKE observation upon varying the laser pulse duration it was possible to draw a complete state diagram of pulse duration versus fluence, as reported in Fig.~\ref{fig:stack}d. The type of reversal, either concentric rings or single domain, is dependent on the multilayer respective thicknesses. As shown in Fig.~\ref{fig:stack}c for positions P1, P2 and P3, the threshold fluences TF2 and TF3 defining the presence of the second ring, having the same magnetization as the initial state (red contrast), are brought closer with increasing Tb and Co thickness, merging together for Tb:0.86nm / Co:1.45nm at P3, to form a single reversed domain. Our model links this effect to an increase in the Gilbert damping due to the increased Tb content. 
As reported in \cite{PEN23} and contrary to Gd based materials, the fluence required to reverse and stabilize a given number of rings has little or no dependence on the laser pulse duration. Single pulse reversals were obtained for pulse duration from 50fs up to 12ps, the maximum pulse duration available on the laser setup. 

Recently we have studied the possible origin of these peculiar AOS properties pointing towards an in-plane magnetization reorientation with a precessional switching mechanism. The reported state diagrams observed for Tb/Fe, and Tb$_{32}$Co$_{68}$/Co trilayer \cite{ PEN23} share the same similarities as those of the [Tb/Co] multilayers, proving that a similar magnetization reversal process takes place in these Tb based multilayer systems.  

The independence on pulse duration suggests that the reversal mechanism proceeds for a time longer than the laser pulse itself and is slower than the one observed in Gd based samples \cite{RAD11}. Considering a two temperature model, at the short time scale up to a few ps the behaviour of the electron temperature $T_{e}$ is highly dependent on the laser pulse duration. Within few ps the $T_{e}$ relaxes to the phonon temperature $T_{p}$ and, from there on, the time evolution of the two becomes indistinguishable. For longer times, in the order of hundreds of ps, the reached temperatures are only due to the laser fluence and cooling of the system to the thermal bath. Based on the experimental results, we assume that the key dynamics for the reversal occur during the longer time phase. 
This interpretation is based on the fact that super-diffusive spin currents \cite{choi2014spin}, hot electrons and ultra-fast demagnetization \cite{beaurepaire1996ultrafast} are effects highly dependent on pulse duration. 
This is contrary to our experimental observations where the final reversed state is essentially independent of the laser pulse length up 12ps and probably beyond. The initial fast response phase certainly plays a role to start the reversal, and may be more important for fs pulses, but does not determine the end-state. 

Hence, we developed a macrospin model based on the Landau-Lifshitz-Gilbert equation coupled to two temperature model to assess whether the above assumptions are consistent with the main features of our observations, providing an insight into the underlying reversal process taking place in at the ps timescale. 
\newpage
\marginpar{
\vspace{.7cm} 
\color{Gray} 
\textbf{Figure \ref{fig:LLG}} 
a) Magnetization dynamics 3D view as predicted by macrospin model. b) M$_{z}$ and Q factor time traces for fluences 6.3mJ/cm$^{2}$, 9.0mJ/cm$^{2}$, 18.0mJ/cm$^{2}$, 21.0mJ/cm$^{2}$. c) Expected magnetization pattern after 100fs for a single Gaussian shaped laser pulse with peak fluences as in a). Simulations in a),b) and c) were performed for $\theta_{K}=5^{o}$, $ M_{s}=1.42$MA/m and  $ K_{u}(300K)=1.1 \cdot \frac{ \mu_{0}M_{s}^{2}}{2}  $, $\alpha=0.1$. d) For position P1 (Fig. 1b), experimental background subtracted MOKE images, laser: 50fs and 10.3mJ/cm$^{2}$, 14.4mJ/cm$^{2}$, 16.9mJ/cm$^{2}$, 19.7mJ/cm$^{2}$.
}

The model accounts for two key features: i) fast drop of the factor $ Q = 2 K_{u}/(\mu_{0}M_{s}^{2})$ with increasing temperature and ii) tilt of the uniaxial anisotropy axis with respect to the out-of-plane direction $\theta_{K}$ \textgreater $ 0^{o}$. The dynamics shown in Fig.\ref{fig:LLG} results from the sharp reduction of the Q factor due to the strong dependency of uniaxial anisotropy $K_{u}$ with temperature, the phenomenological relation used is $K_{u}(T) = K_{u0}[1-(\frac{T}{T_{C}})^{a}]^{b}$, where $T_C = 450K$ is a critical temperature, the exponent a=1.73 and b=2 [\cite{sanchez2022real}]. When the Q factor is less than 1 the demagnetizing field overcomes the anisotropy and the magnetization starts precessing towards the plane. The precession time depends on the energy absorbed, i.e. the laser fluence. During the cooling, the Q factor increases again and if the precession endpoint has overcome the energy maximum given by the tilt of the anisotropy the magnetization reverses. As the fluence increases, the number of precessions made by the magnetization increases. Depending whether the number of half precessions is odd or even, the magnetization will end up as reversed or not reversed resulting in the concentric rings pattern shown in Fig.~\ref{fig:stack}c and Fig.~\ref{fig:LLG}d.

Although the model is fairly simple, it is able to correctly predict the main features of our [Tb/Co] multilayer system. The perpendicular anisotropy originating from the Tb/Co interface is known to have an important temperature dependence, with typical blocking temperatures below $200^{o}C$ due to the low Curie temperature of Tb (237K for bulk measurements \cite{Terbium1958}). Simulations were done assuming a constant $M_{s}$, although we expect it to increase with temperature due to the fact that the AO-HIS occurs only in Co rich regions. By considering this increase, the Q factor fast drop will be further helped and thus favoring the reversal at lower fluence. The assumption of a tilted axis for the anisotropy is supported by the hysteresis loop measurements in Fig.\ref{fig:AOSdem}c, showing a gradual saturation for higher perpendicular fields, which indicates the existence of an in-plane component. The presence of in-plane uniaxial anisotropy in [Tb/Co] multilayers has already been reported \cite{ertl1992structure}, providing further justification to consider it. As shown by the magnetization time traces in Fig.\ref{fig:LLG}b, the total time for reversal is expected to be of the order of tens/hundreds of ps, but the final magnetization state is determined by the first few precessions. Our model provides a good qualitative  and consistent understanding of the observed magnetization reversal patterns, even without considering the effects of the initial pulse heating phase that require atomistic or Landau-Lifshitz-Bloch (LLB) models which will be addressed in a future study.

\begin{wrapfigure}[31]{l}{66mm}

\includegraphics[width=65mm]{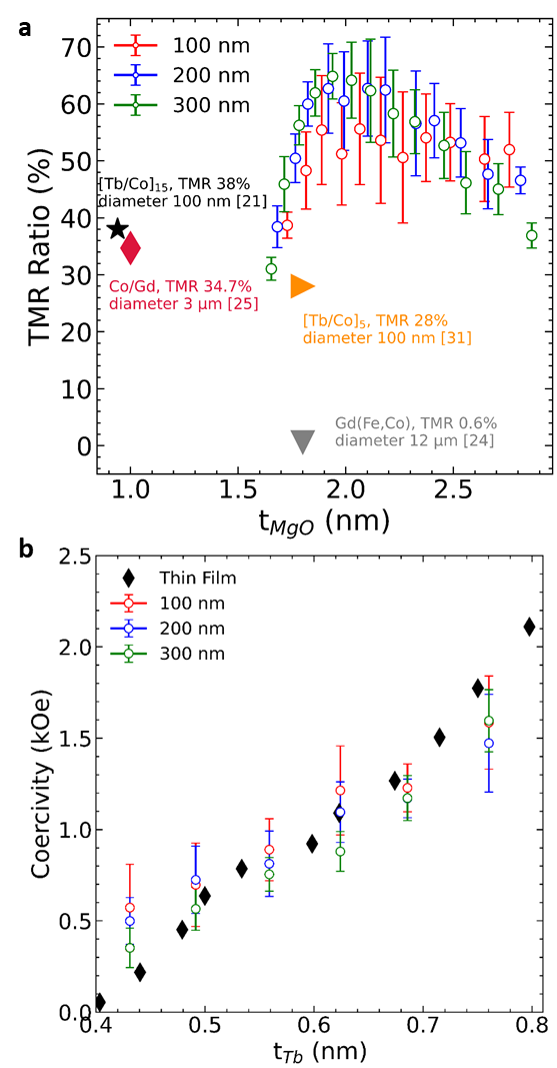}
\captionsetup{labelformat=empty} 
\caption{} 
\label{fig:TMR} 
\end{wrapfigure}

\begin{wrapfigure}[19]{l}{75mm}
\includegraphics[width=75mm]{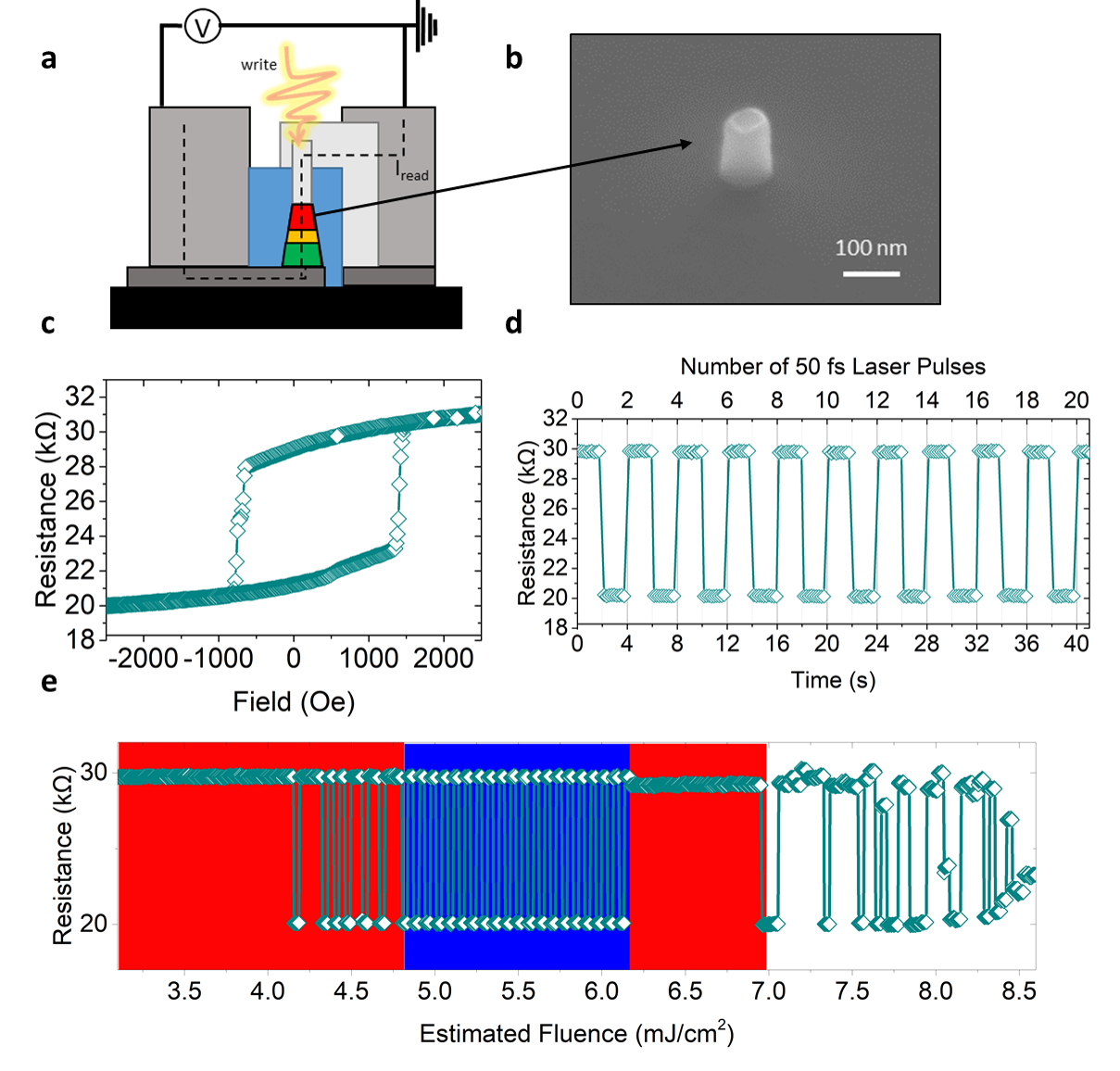}
\captionsetup{labelformat=empty} 
\caption{} 
\label{fig:AOSdem} 
\end{wrapfigure}

\newpage
\marginpar{
\vspace{.7cm} 
\color{Gray} 
\textbf{Figure \ref{fig:TMR}} 
 a) The device average TMR versus naturally oxidized MgO thickness for nominal diameter between 100nm to 300nm and thicknesses  t$_{Tb}$=0.5nm and t$_{Co}$=1.4nm, respectively compared to literature values\cite{AVI20, WAN22, CHE17, OLI20}. b) Coercive field vs Tb thickness comparison between thin film sample and nano-patterned devices.
 }

Next, we introduced this Tb/Co based thin  films as storage layer in full MTJ stacks with optical access enabling AOS writing and electrical readout.
Nanopillars were fabricated on a cross-wedge thickness sample annealed at $250^{o}C$, keeping a Co layer thickness fixed at 1.4nm while varying the Tb layer thickness from 0.4nm to 1.1nm across a 100mm Si(100) wafer, to include the known AOS thickness region for Tb ~0.7-0.9nm. Perpendicularly, a cross-wedge of the MgO barrier was deposited, with thickness ranging from 1.1nm to 3.0nm to characterize the optimal Mg natural oxidation thickness to achieve the highest TMR ratio. 

The nanofabrication process includes an indium tin oxide (ITO) as transparent conductive electrode to provide an optical access to the top electrode of magnetic tunnel junction (details in [\cite{OLI20}]). Fabricated MTJ devices with diameters from 300nm down to 80nm show no impact on the expected AOS stack properties from the reference layer with typical offset fields below 200Oe. Indeed, high perpendicular anisotropy at film level is maintained after nanopatterning as shown by the patterned and thin film coercivity in Fig.~\ref{fig:TMR}b. The coercivity depends on Tb/Co thickness, but stays comparable before and after patterning without substantial dependency on diameter. Device average TMR for each pillar nominal diameters from 80nm to 300nm versus naturally oxidized MgO thickness are compared to literature values in Fig.~\ref{fig:TMR}a, for a Tb thickness of 0.5nm. 
The natural oxidation time was fixed to 2 steps of 10s at 150mbar. The optimal Mg thickness is 2nm, below this value the tunnel barrier is over oxidized leading to lower TMR values. This stable reference layer combined with optimized Mg natural oxidation thickness allows us to get highest TMR values of 74\% representing a 2 times improvement compared to the highest AOS-MTJ reported so far\cite{ CHE17, AVI19, AVI20, OLI20, WAN22}. 


The all-optical switching properties of nanosized MTJ pillars were investigated using linearly polarized fs laser pulses while measuring the resistance of the junction, as illustrated in the schematics of Fig.~\ref{fig:AOSdem}a. The electrical readout was performed using a digital multimeter connected to the device top and bottom electrodes by wire-bonding. The resistance was measured applying a continuous reading voltage of 10mV. All measurements were performed without any external magnetic field, without compensating offset fields from the reference layer. Fig.~\ref{fig:AOSdem}d shows an example of a 100nm AOS-MTJ that can be repeatedly toggled between parallel (P) and anti-parallel (AP) states, showing the full 50\% TMR resistance change expected from the R-H loop in Fig.~\ref{fig:AOSdem}c. 
The upper limit for absorbed energy, calculated by neglecting reflection and thus considering all laser fluence multiplied by the area of the pillar, was estimated to be 392fJ/bit. This is already much lower than the dissipation writing energy of existing technologies like hard disk drives (10-100nJ)\cite{hylick2008analysis}, FLASH-memory (10nJ)\cite{liu2009nanoscale}, and comparable to state-of-art STT-RAM (450pJ-100fJ)\cite{wang2013low, KIM19}.

\marginpar{
\vspace{.7cm} 
\color{Gray} 
\textbf{Figure \ref{fig:LLG}} 
\textbf{ a) Schematic illustration of all-optical writing and electrical reading of AOS-MTJ. b) SEM image of 100nm pillar. c) AOS-MTJ hysteresis  R-H loop, field applied out-of-plane. d) TMR measurement as a function of time upon laser-pulse excitation. e) TMR measurement using constant 1Hz laser frequency and incremental laser power for each pulse.}
}

In Fig.~\ref{fig:AOSdem}e we compare the fluence write levels between the patterned tunnel junction and the equivalent AOS stack at thin film level. The switching dependence on fluence appears to be preserved after patterning. In the performed experiment, the laser was shot repeatedly at a frequency of 1Hz, while increasing the laser power with a step of 0.01mW every second. It allows probing the minimum fluence for which a reversal is obtained. At low fluence, there is no switching, and some switching events are observed starting at $4mJ/cm^2$. In a region from ~5-6$mJ/cm^2$ the switching probability is 100\% as highlighted in blue in Fig.~\ref{fig:AOSdem}. For even higher fluence, in the region highlighted in red, switching is no longer observed, followed by random switching before an eventual device degradation. There is a strong correlation between this established device phase diagram and that of the thin film from Fig.~\ref{fig:stack}d), indicative of a very similar reversal mechanism responsible for the alternating of reversal and non-reversal regions. In the framework of our model the same precessional switching reversal process determines the AOS properties of the patterned device, despite the fact that some stray field from the reference layer is acting on the storage layer.


\section{\label{sec:level3}Conclusion:}

In conclusion, we have integrated a bottom reference p-MTJ with an all-optical-switchable [Tb/Co]$_{x5}$ based storage layer, the resulting  continuous film and patterned devices having similar magnetic properties. This system shows all-optical-switching for Co-rich regions close to the magnetization compensation region at room temperature for a Tb thickness range of 0.6nm to 0.9nm, and Co 1.2nm to 1.5nm. We developed a macrospin model able to predict several characteristic features observed experimentally. Thus the magnetization reversal process appears to be driven by an in-plane reorientation of the magnetization coupled to a precessional switching mechanism. The model explains the experimentally observed concentric rings of opposite magnetization directions that appear at high laser fluence. Remarkably, both experiments and simulations indicate that the threshold fluences do not depend on the pulse duration but rather on the AOS layer thicknesses. This represents a significant advantage of the Tb/Co material providing high resilience to pulse length variability at the device level. Furthermore, the implementation of a stable reference layer and optimization of the MgO natural oxidation allowed for TMR values of 74\%, the highest reported so far for AOS-MTJs. Field-free helicity independent all optical toggle switching was demonstrated on 100nm patterned [Tb/Co] p-MTJ devices, driven by 50fs laser pulses for an absorbed energy of 392fJ/bit. These findings pave the way towards nanoscale devices for optospintronic embedded memories combining non-volatily with ultra-fast and energy-efficient writing.

\section*{Acknowledgments}
We acknowledge financial support from the ANR (ANR-17-CE24-0007 UFO project) and European Union’s Horizon 2020 research and innovation programme under Marie Skłodowska-Curie grant agreement No 861300 (COMRAD), this work is supported by the Institute Carnot ICEEL for the project “CAPMAT” and FASTNESS, the Région Grand Est, the Metropole Grand Nancy, for the Chaire PLUS by the impact project LUE-N4S, part of the French PIA project “Lorraine Université d’Excellence” reference ANR-15-IDEX-04-LUE, the “FEDERFSE Lorraine et Massif Vosges 2014-2020” for PLUS and IOMA a European Union Program.


\bibliography{AOS}

\bibliographystyle{abbrv}

\end{document}